\documentclass[12pt]{article}
\usepackage{scicite}
\usepackage{times}
\usepackage{amsmath,amssymb}
\usepackage{graphicx}
\usepackage{color}
\usepackage{color}
\usepackage{changes}
\usepackage{tikz}
\usepackage{bm}
\usepackage{enumitem}
\usepackage{lipsum}
\usepackage{changes}
\usepackage{microtype}

\usepackage{hyperref}

\definecolor{webgreen}{rgb}{0,.35,0}
\definecolor{webbrown}{rgb}{.6,0,0}
\definecolor{RoyalBlue}{rgb}{0,0,0.9}
\definecolor{purp}{rgb}{0.6,0.05,0.8}
\definecolor{ora}{rgb}{0.7,0.35,0.02}

\hypersetup{
   colorlinks=true, linktocpage=true, pdfstartpage=3, pdfstartview=FitV,
   breaklinks=true, pdfpagemode=UseNone, pageanchor=true, pdfpagemode=UseOutlines,
   plainpages=false, bookmarksnumbered, bookmarksopen=true, bookmarksopenlevel=1,
   hypertexnames=true, pdfhighlight=/O,
   urlcolor=webbrown, linkcolor=RoyalBlue, citecolor=webgreen,
   pdfauthor={Siheng Chen, Gary P. T. Choi, L. Mahadevan},
   pdfsubject={Deterministic and stochastic control of kirigami topology}
}

\newenvironment{sciabstract}{%
\begin{quote} \bf}
{\end{quote}}

\newcounter{lastnote}

\newcommand*\circled[1]{\tikz[baseline=(char.base)]{
            \node[shape=circle,draw,inner sep=0.85pt] (char) {#1};}}

\newtheorem{theorem}{Theorem}

\begin{document}

\author{Siheng Chen$^{1\dagger}$, Gary P. T. Choi$^{1\dagger}$, L. Mahadevan$^{1,2,3\ast}$\\
\\
\footnotesize{$^{\dagger}$These authors contributed equally to this work.}\\
\footnotesize{$^{1}$John A. Paulson School of Engineering and Applied Sciences, Harvard University}\\
\footnotesize{$^{2}$Departments of Physics, and Organismic and Evolutionary Biology, Harvard University}\\
\footnotesize{$^3$Kavli Institute for Nanobio Science and Technology, Harvard University, Cambridge, MA 02138, USA}\\
\footnotesize{$^\ast$To whom correspondence should be addressed; E-mail: lmahadev@g.harvard.edu}
}
\title{Deterministic and stochastic control of kirigami topology}
\date{} 

\baselineskip24pt

\maketitle

\begin{sciabstract}
Kirigami, the creative art of paper cutting, is a promising paradigm for mechanical metamaterials. However, to make kirigami-inspired structures a reality requires controlling the topology of kirigami to achieve connectivity and rigidity. We address this question by deriving the maximum number of cuts (minimum number of links) that still allow us to preserve global rigidity and connectivity of the kirigami. A deterministic hierarchical construction method yields an efficient topological way to control both the number of connected pieces and the total degrees of freedom. A statistical approach to the control of rigidity and connectivity in kirigami with random cuts complements the deterministic pathway, and shows that both the number of connected pieces and the degrees of freedom show percolation transitions as a function of the density of cuts (links). Together this provides a general framework for the control of rigidity and connectivity in planar kirigami.
\end{sciabstract}

\newpage
Kirigami, the traditional art of paper cutting, has inspired the design of a new class of metamaterials with novel shapes~\cite{Grima00,Grima04,Grima05,Shan15,Rafsanjani16}, electronic/ mechanical properties~\cite{Kane14,Sun12,Rafsanjani17,Blees15,Isobe16,Tang17}, and auxetic behavior~\cite{Mitschke13,Kolken17,Gatt15}. In these studies, typically the geometry (the shape of the deployed kirigami) and the topology of cuts (spatial distributions of the cuts) are prescribed. Recently we have shown how to modulate the geometry of the kirigami structures~\cite{Choi19} by varying the size, and orientation of the cuts to solve the inverse problem of designing kirigami tessellations that can be deployed to approximate given two and three dimensional shapes. However, in these and other studies, the topology of the cuts is not a variable that can be changed to achieve a prescribed shape or mechanical response. Here, we relax this constraint and allow the topology of the kirigami structures to be a design variable that we can control either in a deterministic or a stochastic way to achieve a given connectivity or rigidity in aperiodic planar structures. 

\begin{figure}[!t]
\centering
\includegraphics[width=0.8\textwidth]{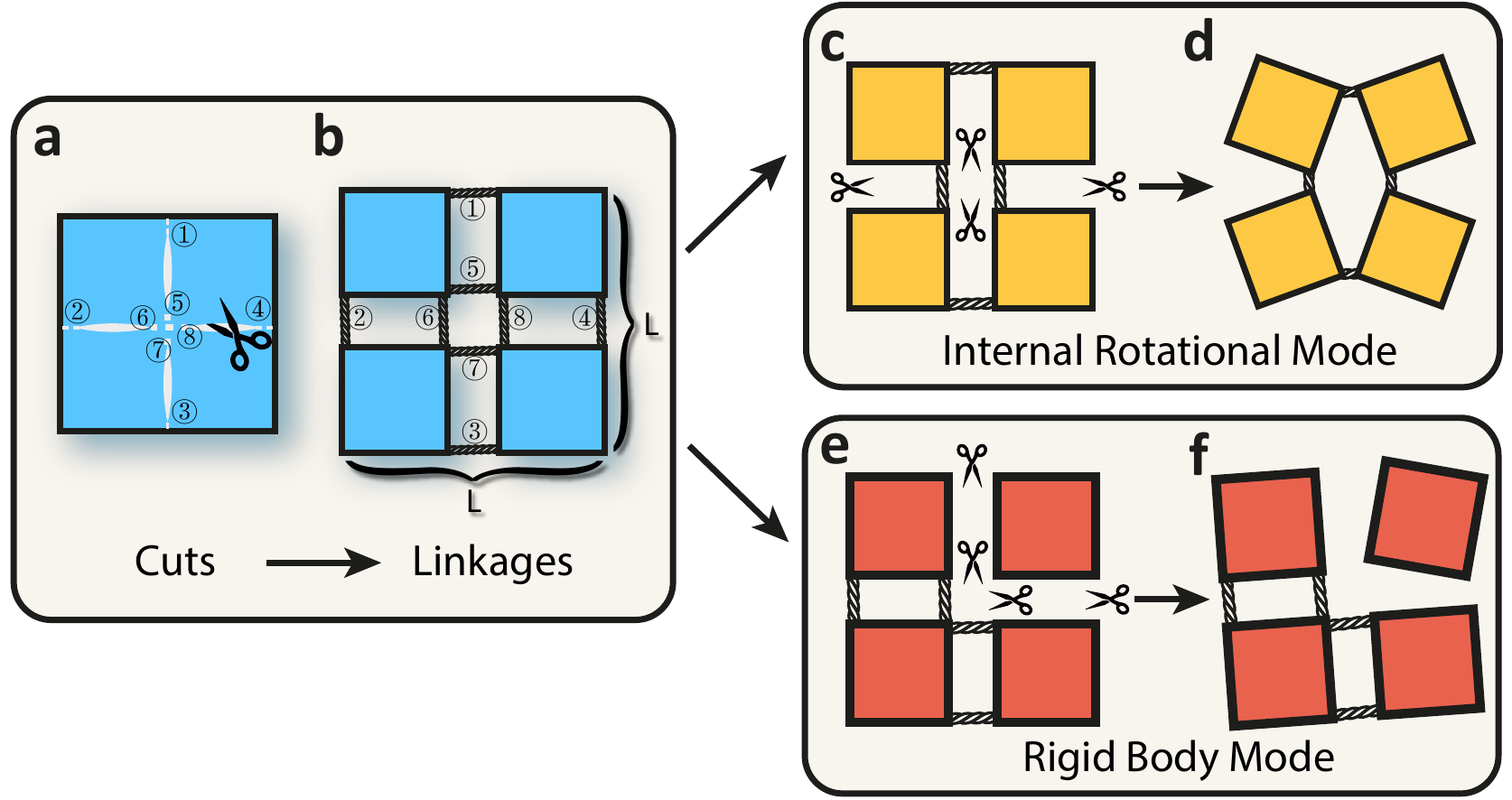}
\caption{Quad kirigami and two types of floppy modes. {\bf a)} The cuts are along edges of quads except at the vertices, so that the pattern is equivalent to a linkage shown in {\bf b)}. Removing certain links can increase the DoF of the structure, adding some internal rotational mechanisms ({\bf c}-{\bf d}), or increase the number of connected components (NCC), adding translational and rotational rigid body modes ({\bf e}-{\bf f}).}
\label{fig:cut_link}
\end{figure}

We start by considering cutting a thin sheet of material (width$=$height$=l$) to obtain a planar kirigami system, i.e. an initial structure that has cuts that allow it to be transformed into a new shape using local rotations with no energy cost. Since cuts at random locations with random directions and random lengths are unlikely to lead to a system that can respond via purely rotational modes, we start with a quad kirigami structure~\cite{Grima00}, a simple four-fold symmetric auxetic structure as shown in Fig.~\ref{fig:cut_link}{\bf a}. To constrain the infinite phase space of cuts to a finite combination, we adopt the following assumptions: (1) We only allow horizontal and vertical cuts on grid lines with equal spacing $d$. There are $L=l/d$ horizontal and $L$ vertical grid lines. (2) The sheet is cut along all the grid lines, except in the vicinity of vertices where the grid lines intersect (Fig.~\ref{fig:cut_link}{\bf a}). (3) Since kirigami is deployable, the quads need to rotate around their corner hinges and we assume that the width of hinges is infinitesimally small. Around internal vertices, there are four infinitesimally small segments which we can independently decide to cut or not, as illustrated by the dashed lines marked as \circled{5}-\circled{8} in Fig.~\ref{fig:cut_link}{\bf a}. For boundary vertices, there is only one segment (\circled{1}-\circled{4} in Fig.~\ref{fig:cut_link}{\bf a}). (4) The quads connected by the corner hinges can rotate freely without any energy. These assumptions and simplifications allow us to transform the cutting problem into a linkage problem, as shown in Fig.~\ref{fig:cut_link}{\bf b}, where all the quads are separated, and each pair of neighboring corners (nodes) of quads can be connected via a link. This is a one-to-one mapping, as each link in Fig.~\ref{fig:cut_link}{\bf b} has a unique corresponding segment for cutting in Fig.~\ref{fig:cut_link}{\bf a}. Note that the links do not have actual length - they only force the corners of two quads to be connected (with the same spatial coordinates). With this setup, the problem of cutting to derive a kirigami structure is converted to choosing certain number of links among the $4(L-1)^2+4(L-1) =4L(L-1)$ links to be connected. By adding and removing a specific set of links, one can manipulate the number of degrees of freedom (DoF) and number of connected components (NCC) in the system. Moreover, one can even control the types of floppy modes associated with internal rotational mechanisms (Fig.~\ref{fig:cut_link}{\bf c}-{\bf d}) and rigid body modes coming from additional pieces (connected components) (Fig.~\ref{fig:cut_link}{\bf e}-{\bf f}). 
We note that while we focus on the quad kirigami as it is the simplest periodic planar structure, all our results also hold for kagome kirigami, a six-fold auxetic structure formed by equilateral triangles (see SI Appendix, section S5). 

Having shown the equivalence between connectivity and rigidity of kirigami to that in a linkage, we ask how we might vary the number and spatial distribution of the links to control the number of connected pieces (connectivity) and floppy modes (rigidity) in a kirigamized sheet? We show that prescribing the microscopic cuts in a kirigami tessellation using a hierarchical linkage pattern yields any targeted rigidity and connectivity. Furthermore, in the absence of microscopic control, if we are still able to control just the coarse-grained density of cuts, we show the existence of percolation transitions that allow for the control of both connectivity and mechanical properties in an exquisitely sensitive way, similar in some aspects to rigidity percolation in both planar networks and origami~\cite{Jacobs95,Jacobs97,ellenbroek2011rigidity,Lubbers19,Chen19}. (Here we note that the ``rigidity'' represents the DoF (or number of floppy modes) introduced above, which is different from the ``rigidity percolation'' transition that describes a global change in the network, e.g. Fig.~\ref{fig:connectivity_percolation}.)

The connectivity and rigidity of a kirigami tessellation is controlled by the geometrical constraints associated with preserving the four edge lengths and one diagonal length on the four vertices of each quad in a unit cell shown in Fig.~\ref{fig:cut_link}. These can be written in the form
\begin{align}
g_{\text{edge}}(\bm{x_i},\bm{x_j})=||\bm{x_i}-\bm{x_j}||^2-d^2=0, 
\end{align} 
where node $i$ and node $j$ are connected by an edge in a quad, $i,j \in \{1, ..., 4L^2\}$. Additionally, the introduction of rigid links at coincident vertices (see Fig.~\ref{fig:cut_link}) implies that there are constraints on the coordinates of the two vertices for each link connecting node $i$ and node $j$ written as:
\begin{align}
&g_{\text{link}_x}(\bm{x_i},\bm{x_j})=x_{i_1}-x_{j_1} =0,\\
&g_{\text{link}_y}(\bm{x_i},\bm{x_j})=x_{i_2}-x_{j_2} =0,
\end{align}
where $\bm{x_i} = (x_{i_1}, x_{i_2})$ and $\bm{x_j} = (x_{j_1}, x_{j_2})$. These set of constraints determine the range of motions associated with infinitesimal rigidity in terms of the rigidity matrix $\bm{A}$ where $A_{ij}=\partial g_i/\partial x_j$, so that the DoF is related to the rank of $\bm{A}$ via the relation~\cite{Guest06,Lubensky15} (see SI Appendix, section S1):
\begin{align} \label{eqt:dof_rank}
\text{DoF}=8L^2-\textnormal{rank}(A).
\end{align}

\section*{Kirigami with prescribed cuts}
\subsection*{Minimum rigidifying link patterns}

To address the question of rigidity control with prescribed cuts, we note that the decrease in the total DoF by adding one link is either $0$, $1$, or $2$ (see SI Appendix, section S1). From a mathematical perspective, each link adds two rows to the rigidity matrix $A$, and the rank of $A$ can increase by 1, 2, or remain unchanged. Therefore, we can calculate the minimum number of links (denoted as $\delta(L)$) required for \emph{rigidifying} an $L\times L$ kirigami - having no extra DoF besides the rigid body motions. Define a \emph{minimum rigidifying link pattern }(MRP) to be a link pattern (a set of positions for links) that rigidifies the kirigami with exactly $\delta(L)$ links. Note that there are $3L^2$ DoF if all the links are disconnected, and there are 3 DoF if all the links are connected. Since each link reduces the DoF by at most 2, $\delta(L)$ links can at most reduce $2\delta(L)$ DoF and hence $3L^2-2\delta(L) \leq 3$. Therefore,
$
\delta(L) \geq \left\lceil \frac{3L^2 - 3}{2} \right\rceil,
$
where $\left \lceil \cdot \right \rceil$ is the ceiling function rounding up the number to the nearest integer. Note that $\delta(L)$ might be greater than the lower bound $\left\lceil \frac{3L^2 - 3}{2} \right\rceil$, since there might be no way that the $\left\lceil \frac{3L^2 - 3}{2} \right\rceil$ links added are all non-redundant. It is natural to ask whether this lower bound for $\delta(L)$ is achievable (optimal), and furthermore, is it always possible to find a rigidifying link pattern with exactly $\left\lceil \frac{3L^2 - 3}{2} \right\rceil$ links?

We give a constructive proof of the optimality of the above lower bound by developing a \emph{hierarchical construction} method for constructing MRPs for any system size $L$, where we combine the patterns for small $L$ to construct the patterns for large $L$. First, we show that if the lower bound is achievable for two odd numbers $L = L_1$ and $L = L_2$ (i.e. $\delta(L_1) =\left\lceil \frac{3L_1^2 - 3}{2} \right\rceil$ and $\delta(L_2) =\left\lceil \frac{3L_2^2 - 3}{2} \right\rceil$), it is also achievable for $L=L_1L_2$. The key idea is that, if we treat an $L_1L_2 \times L_1L_2$ kirigami as $L_2\times L_2$ large blocks of $L_1\times L_1$ quads, we can rigidify each large block of $L_1\times L_1$ quads by an MRP for $L = L_1$, and then link and rigidify all the $L_2\times L_2$ large rigid blocks by an MRP for $L= L_2$. This gives a rigidifying link pattern for the $L_1L_2 \times L_1L_2$ kirigami, with the total number of links being
\begin{equation}
\begin{split}
&L_2^2\delta(L_1)+\delta(L_2) = L_2^2\left \lceil \frac{3L_1^2 - 3}{2} \right\rceil+\left \lceil \frac{3L_2^2 - 3}{2} \right\rceil\\
&=L_2^2\frac{3L_1^2 - 3}{2}+\frac{3L_2^2 - 3}{2}=\frac{3L_1^2L_2^2 - 3}{2} =\left \lceil \frac{3(L_1 L_2)^2 - 3}{2} \right\rceil.
\end{split}
\end{equation}
Therefore $\delta(L_1 L_2) = \left \lceil \frac{3(L_1 L_2)^2 - 3}{2} \right\rceil$, and hence such a link pattern is an MRP for $L = L_1 L_2$. Similarly, if $L_1$ is odd and $L_2$ is even, we can show that this relationship still holds (see SI Appendix, section S1). In fact, using the hierarchical construction method and MRPs for small $L$ (with the optimality of the lower bound for those small $L$ cases proved using the rigidity matrix computation), we can prove that the lower bound $\delta(L)$ is optimal for all $L$.

\begin{theorem} \label{thm:rigidity_min}
For all positive integer $L$, 
$
\delta(L) = \left\lceil \frac{3L^2 - 3}{2} \right\rceil.
$
\end{theorem}
{\it Proof.} We outline the proof here and leave the details in SI Appendix, Theorem S1-S3. Our idea is to use perfectly periodic, prime partitions to decompose an $L\times L$ kirigami into blocks of kirigami with smaller size.

\begin{figure}[t]
\centering
\includegraphics[width=0.9\textwidth]{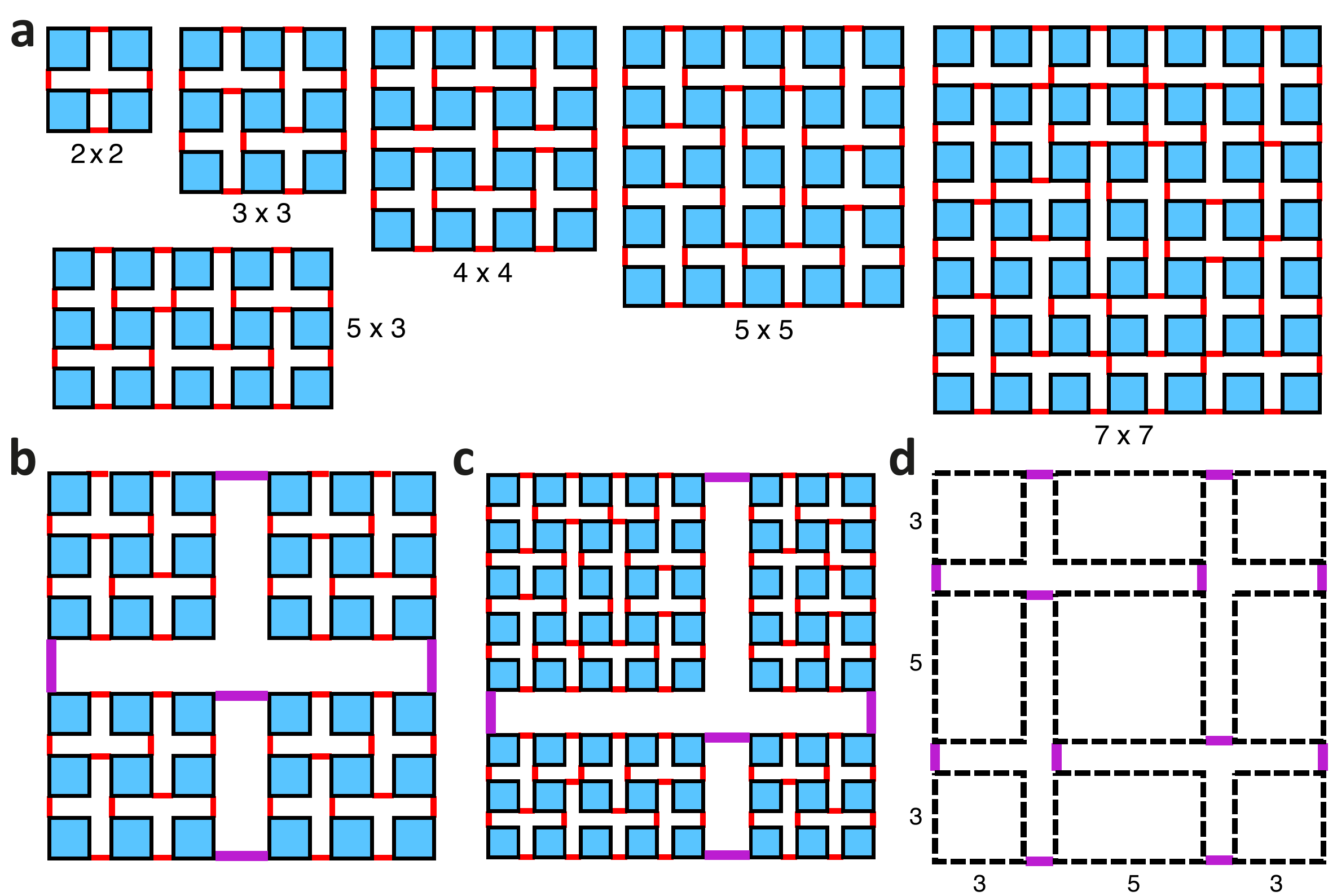}
\caption{The construction of minimum rigidifying link patterns (MRPs). {\bf a)} Explicit construction of MRPs for $L=2,3,4,5,7$, and for $3\times 5$. {\bf b)} The construction of an MRP for $L = 6$ using hierarchical construction. The $6 \times 6$ kirigami is decomposed into four large blocks of size $3\times 3$. Each large block is rigidified using an MRP for $L = 3$, and then the four large rigid blocks are linked and rigidified using an MRP for $L=2$. {\bf c)} The construction of an MRP for $L = 2^k, k \geq 3$ using hierarchical construction. A $2^3 \times 2^3$ kirigami is decomposed into large blocks of size $3\times 3$, $5\times 5$ and $3\times 5$. Each large block is rigidified using an MRP, and then the four large rigid blocks are linked and rigidified using an MRP for $L=2$. {\bf d)} The construction of an MRP for odd primes $p \geq 11$ using hierarchical construction. A $11 \times 11$ kirigami is decomposed into large blocks of size $3\times 3$, $5\times 5$ and $3\times 5$. Each large block is rigidified using an MRP, and then the four large rigid blocks are linked and rigidified using an MRP for $L=3$. }
\label{fig:minimum_rigidifying_patterns}
\end{figure}

In Fig.~\ref{fig:minimum_rigidifying_patterns}{\bf a}, we explicitly construct MRPs with exactly $\left\lceil \frac{3L^2 - 3}{2} \right\rceil$ links for $L = 2,3,4,5,7$, and an auxiliary MRP for a $3\times 5$ kirigami (see SI Appendix, section S1). The rigidity of these patterns is verified using the rigidity matrix computation \eqref{eqt:dof_rank}. Then, with these small patterns, we can use the hierarchical construction to obtain an MRP for any $L = 2^k \prod p_i^{n_i}$, where $k=0,1,2$ and $p_i$ are odd primes satisfying $\delta(p_i) = \left\lceil \frac{3p_i^2 - 3}{2} \right\rceil$ (see Fig.~\ref{fig:minimum_rigidifying_patterns}{\bf b} and SI Appendix, Theorem S1).

We further design methods to construct MRPs for all $L$ that is a power of 2 (see Fig.~\ref{fig:minimum_rigidifying_patterns}{\bf c} and SI Appendix, Theorem S2) and for all odd primes $p\geq 11$ (see Fig.~\ref{fig:minimum_rigidifying_patterns}{\bf d} and SI Appendix, Theorem S3) by integer partition and hierarchical construction. Ultimately, we can remove all conditions on $L$ and conclude that the $\delta(L) = \left\lceil \frac{3L^2 - 3}{2} \right\rceil$ for all $L$. \hfill$\blacksquare$

Theorem \ref{thm:rigidity_min} provides the most efficient way to rigidify quad kirigami: placing links according to the MRPs (see SI Appendix, section S1 for a flowchart and the algorithmic procedure of the hierarchical construction method). Moreover, for odd $L$, since every link in an $L \times L$ MRP decreases the DoF of the kirigami system by exactly 2, we can obtain a kirigami with $\text{DoF} = 2k+3$ by removing exactly $k$ links from an MRP. By adding a link which reduces the DoF by 1 to such a kirigami system, we can obtain a kirigami with $\text{DoF} = 2k+2$. For even $L$, all but one links in an $L \times L$ MRP reduce the DoF of the system by 2 (except one that reduces the DoF by 1). By removing $k$ links from an MRP, we can again obtain a kirigami with $\text{DoF} = 2k+3$ or $2k+2$. Therefore, any given DoF is achievable. 

\subsection*{Minimum connecting link patterns}
For the connectivity of kirigami, a similar question arises: What is the minimum number of prescribed links for making an $L \times L$ kirigami connected? Obviously, when one link is added, the NCC decreases by either $0$ or $1$. Define $\gamma(L)$ as the minimum number of links required for connecting an $L \times L$ kirigami, and a \emph{minimum connecting link pattern }(MCP) to be a link pattern with exactly $\gamma(L)$ links which connects the $L \times L$ kirigami. Note that there are $L^2$ connected components if all the links are disconnected, and 1 connected component if all the links are connected. Following the same argument as the section above, we have $L^2 - \gamma(L) \leq 1$, thus $\gamma(L) \geq L^2 - 1.$
It turns out that by explicit construction we are able to show that this lower bound is optimal for all $L$.
\begin{theorem}
\label{thm:connectedness_min}
For all positive integer $L$, $\gamma(L) = L^2 -1.$
\end{theorem}
(See SI Appendix, section S2 for the detailed proof.) Similar to the construction of MRPs, the construction of MCPs can be done by hierarchical construction. Besides, since every link in an MCP decreases the NCC by exactly 1, we can obtain a kirigami with $k+1$ connected components by removing $k$ links from an MCP. Therefore, Theorem \ref{thm:connectedness_min} provides us with an efficient way for constructing a kirigami system with any given NCC. 

From Theorem \ref{thm:rigidity_min} and \ref{thm:connectedness_min}, we can easily see that for $L \geq 2$, 
\begin{equation}
\delta(L) = \left\lceil \frac{3L^2 - 3}{2} \right\rceil > L^2 - 1 = \gamma(L).
\end{equation}
This implies that there is no MRP which is also an MCP for $L \geq 2$. It also suggests that in general rigidifying a kirigami requires more effort (links) compared to connecting the quads in kirigami. We remark that by using MRPs and MCPs, we can achieve a certain level of simultaneous control of rigidity and connectivity. For instance, we can achieve an $L \times L$ system with NCC $= d^2$ and DoF $= 3d^2$ precisely if $d$ is a factor of $L$ (see SI Appendix, section S4 for a detailed discussion).

\subsection*{Enumeration of minimum link patterns}
It is noteworthy that the constructions of MRPs and MCPs are not necessarily unique. Denote the number of MRPs and MCPs in an $L \times L$ kirigami by $n_r(L)$ and $n_c(L)$ respectively. To obtain $n_r(L)$, note that the total number of links in an $L \times L$ kirigami is $4L(L-1)$ and hence there are ${4L(L-1)\choose \left\lceil \frac{3L^2 - 3}{2} \right\rceil}$ possibilities to examine. As this number grows rapidly with $L$, obtaining $n_r(L)$ by direct enumeration is nearly impossible. Nevertheless, we can make use of the hierarchical construction to obtain a lower bound of $n_r(L)$ (see SI Appendix, section S1 and Table S1). Similarly, for $n_c(L)$, there are ${4L(L-1)\choose L^2-1 }$ possibilities to examine, which becomes impossible for enumeration for large $L$. This time, we can make use of the \emph{Kirchhoff's matrix tree theorem}~\cite{Kirchhoff47} to obtain $n_c(L)$ (see SI Appendix, section S2 and Table S2). Comparing $n_r(L)$ and $n_c(L)$, we observe that there are much more MCPs than MRPs, suggesting that it is much easier to connect a kirigami than to rigidify a kirigami.

Note that both $n_r(L)/{4L(L-1)\choose \left\lceil \frac{3L^2 - 3}{2} \right\rceil}$ and $n_c(L)/{4L(L-1)\choose L^2-1 }$ become extremely small as $L$ increases (see SI Appendix, Table S1 and S2). This suggests that for large $L$, it is almost impossible to obtain an MRP (or MCP) by randomly picking $\left\lceil \frac{3L^2 - 3}{2} \right\rceil$ (or $L^2-1$) links. Thus, our hierarchical construction is a powerful tool for rigidifying or connecting a large kirigami, and for further achieving a given DoF or NCC, with the minimum amount of materials (links). 

\section*{Kirigami with random cuts}
Controlling the link patterns (locations) according to MRPs and MCPs is a powerful tool to control rigidity and connectivity, but requires exquisite control of every link at the microscopic level, which is difficult to achieve. A natural question that then arises is what if we cannot control the microscopic details of the link patterns? Could we still achieve connectivity and rigidity control by only tuning the fraction of randomly added links in the linkage graph?

\subsection*{Connectivity percolation}
We start by defining the link density for an $L \times L$ kirigami to be $\rho(L) = c/c_{\text{max}}$, where $c$ is the number of randomly added links and $c_{\text{max}}=4L(L-1)$ is the maximum number of links. Furthermore, we state that two quads are connected as long as one of the links between them is present, and a connected component as a set of quads among which every two of the quads are connected by a series of links. To study the NCC in quad kirigami with $c$ random links (or equivalently, $(c_{\text{max}}-c)$ random cuts), we denote the NCC in the $L \times L$ kirigami by $T$, noting that when $c=0$, $T=T_{\text{max}}=L^2$, while when $c=c_{\text{max}}$, $T=T_{\text{min}}=1$. 

To understand what happens at intermediate values of the density, we randomly generate link patterns and calculate $T(\rho)$ (Fig.~\ref{fig:connectivity_percolation}{\bf a}-{\bf c}). Initially, each of the randomly added links simply connects two quads, so $T$ decreases by one as each link is added, i.e. linearly (Fig.~\ref{fig:connectivity_percolation}{\bf d}). Eventually though, they are more likely to be added within one connected component rather than between two connected components, as there are more possible positions for adding links within each connected component, and indeed the decrease of DoF becomes sub-linear (Fig.~\ref{fig:connectivity_percolation}{\bf d}). If we rescale $T$ by $T_{\text{max}}$, we see an exponential decay starting at $\rho=0.3$, finally approaching $1/L^2$ (Fig.~\ref{fig:connectivity_percolation}{\bf d} right inset). We also see that the linear-sublinear transition is universal regardless of the system size $L$ (Fig.~\ref{fig:connectivity_percolation}{\bf d} left inset). 

Furthermore, we notice that at the onset of the exponential decay region, the size of the largest connected component (denoted by $N$) shows percolation behavior. If we rescale $N$ by $N_{\text{max}}=L^2$, we see that $N/L^2$ switches from $0$ to $1$ as the largest connected component suddenly becomes dominant (Fig.~\ref{fig:connectivity_percolation}{\bf a}-{\bf b}, orange quads), and the transition becomes sharper with increasing system size $L$. Numerically, using the curve for $L=100$, we find that the critical transition point is at $\rho_c^* \approx 0.298$ (Fig.~\ref{fig:connectivity_percolation}{\bf f}).

We can also calculate this transition density $\rho_c$ analytically using the dual lattice method~\cite{Stauffer94}. In fact, this $\rho_c$ is the same as the transition density of connectivity percolation of a network, if we consider the kirigami system as a doubly linked network with quads being nodes and links being edges. Since percolation in a random connecting network is a state where there is a connected path from one side to the other, the probability of percolation is equal to the probability of NOT having percolation in the dual lattice (see SI Appendix, section S3). The probability of connecting two quads is $\rho^2+2\rho(1-\rho)$ assuming that each link is equally likely to be connected with probability $\rho$. The probability of these two quads not being connected is thus $(1-\rho)^2$. Denoting the probability of percolation by $P[x]$ where $x$ is the quad connecting probability, we have
$
P[\rho^2+2\rho(1-\rho)]=1-P[(1-\rho)^2].
$
If we let $\rho^2+2\rho(1-\rho)=(1-\rho)^2$, we obtain
$
\rho_c= 1-\frac{1}{\sqrt{2}} = 0.293.
$

Since the percolation density of having a connected path is the same as when the size of the largest cluster becomes dominant, $\rho_c$ coincides with the percolation threshold of $N$, which agrees well with our numerical result $\rho_c^*\approx 0.298$.

\subsection*{DoF with randomly allocated links}
Moving from the connectivity to the modes of motion (DoF) in the structure, we denote the dependence of DoF on link density by $m(\rho)$. These floppy modes can be further classified into two types:
\begin{enumerate}[label=(\alph*)]
\item Internal modes (Fig.~\ref{fig:cut_link}{\bf c}-{\bf d}): The internal rotational mode associated with the movement of some quads with respect to others within one connected component.
\item Rigid-body modes (Fig.~\ref{fig:cut_link}{\bf e}-{\bf f}): The 2 translational motions and 1 rotational motion of each connected component.
\end{enumerate}
Type (a) are the floppy modes that correspond to the nontrivial internal modes of rotation which are hard to find and visualize without using the infinitesimal rigidity approach. Since each connected component has three (type (b)) rigid body motions (3 DoF), the number of internal DoF is equal to the total DoF minus the number of connected components multiplied by three, i.e.

$
m_{\text{int}} = m_{\text{tot}}-3T.
$

With varying $\rho$, the total DoF follows a trend similar to that of the NCC ($T$). When $\rho = 0$, the total DoF is $3L^2$. When a few links are added initially, they are independent constraints, each reducing the DoF by two, and so we see a linear decrease of DoF (Fig.~\ref{fig:connectivity_percolation}{\bf e}). As more links are added, cooperativity between the links sets in and the decrease of DoF becomes sub-linear, and shows exponentially decaying behavior similar to that of $T$: $\log_{10} m\sim -6.0\rho+1.4$ (Fig.~\ref{fig:connectivity_percolation}{\bf e} right inset). Finally, when most links are added, the system becomes rigid, and the DoF approaches 3. If we rescale the DoF by the maximum DoF ($m_{\text{max}}=3L^2$) for each $L$, all of the curves for internal modes collapse (Fig.~\ref{fig:connectivity_percolation}{\bf e} left inset), suggesting that we can control the internal mechanisms in a scale-free manner by simply tuning the link density. 

Given the similarity of our system to 2d random networks~\cite{Jacobs95}, it is perhaps unsurprising to see this percolation behavior. Indeed, if we plot the second derivative of $m$ as a function of constraint number, we can see that this value has a peak between $0.4$ and $0.5$. As the system size becomes larger, it converges to $\rho_r=0.429$ (Fig.~\ref{fig:connectivity_percolation}{\bf g}). This behavior qualitatively agrees with the second-order transition in the generic rigidity percolation in 2D~\cite{Jacobs95}. Quantitatively, an analogy from the MRP/MCP might explain this: since $\delta(L)/\gamma(L)\sim3/2$ when $L$ is large, we expect that $\rho_r = 3 \rho_c/2$ so that $\rho_r =0.439$, which agrees well with our numerical result. 

When all the quads are separated, there are no internal modes, while when all links are connected, the whole system is rigid and there are no internal modes either. Therefore, the number of internal DoF must be non-monotonic as links are added, with a local maximum in the number of internal DoF as a function of $\rho$ (see the schematic plot in Fig.~\ref{fig:connectivity_percolation}{\bf a}-{\bf c}). Indeed, as shown in Fig.~\ref{fig:connectivity_percolation}{\bf e}, the internal DoF first increases as the number of links increases, approaches the number of total DoF, and decreases together with the total DoF as link density further increases. As the density $\rho_i \approx 0.26$, there are maximal number of internal modes, which is different from the density $\rho_c \approx 0.293$, when the whole system first gets almost fully connected. Thus, it is in the neighborhood of $\rho \approx \rho_c$ where we find a range of interesting behaviors as the kirigami structure becomes almost fully connected, while the number of internal mechanisms remains large (similar to the diluted rotating square system as described in a recent paper~\cite{Lubbers19}; See SI Appendix, section S3 for more details). Indeed, from the ratio of the internal DoF and the total DoF $m_{\text{int}}/m_{\text{tot}}$ (Fig.~\ref{fig:connectivity_percolation}{\bf h}), we see that the fraction of internal DoF is large for a range of link density, consistent with the intuition that there are more rotational modes within each connected component of kirigami than rigid body modes of the component.

\subsection*{Constraint redundancy}

We have seen that when links are added to the system, they do one of three things: change the total DoF, change the internal DoF, or simply be redundant. This notion is generic to a number of different systems made of discrete components with constraints; indeed, recently we showed that this is also true for origami~\cite{Chen19}. To address the question here, instead of randomly generating link patterns, we start with no links and add links one by one. At each step, if the DoF of the system does not change after a link is added, we define this link as a \emph{redundant} link. Otherwise, we define it to be \emph{non-redundant}. We can further classify these non-redundant links by examining how they change the internal DoF. Denoting the change of total DoF as $\Delta_t$, the change of rigid body DoF (type (b)) as $\Delta_r$, and the change of internal DoF (type (a)) as $\Delta_i$. It turns out that the only possible combinations of $(\Delta_i, \Delta_r)$ are $(-2, 0), (-1,0), (0,0), (+1,-3)$, the third of which being defined as redundant (see SI Appendix, section S3). In Fig.~\ref{fig:connectivity_percolation}{\bf i}, we show a schematic of the four types of links classified above. Starting from the original configuration (1), there are two connected components, each with three type (b) DoF. The larger component also has two internal DoF. The orange link, after added to the kirigami system, reduces one internal DoF (Fig.~\ref{fig:connectivity_percolation}{\bf i}(2)). The green link connects the two components afterwards, so $\Delta_r=-3$, and $\Delta_i=+1$. There is one more internal mode (Fig.~\ref{fig:connectivity_percolation}{\bf i}(3)). In step (4), the red link freezes the two rotational modes simultaneously, making the whole system rigid. Finally, in (5) and (6), the links added are redundant as they do not change DoF.

We now apply this link-by-link examination to a larger kirigami system with $L=30$. We define the redundancy $r$ in the system as the ratio of the number of redundant links over the number of all unconnected links. Initially, most of the links are non-redundant as they are all independent, and each of them adds one internal mode since they connect two disconnected components (such as Fig.~\ref{fig:connectivity_percolation}{\bf i} (2)-(3)). After the connectivity percolation the links that reduce the internal DoF become dominant, and eventually most additional links become redundant. The fraction of redundant links $n_{\text{Link}}/n_{\text{max}}$ has a transition near $\rho=0.45$ (Fig.~\ref{fig:connectivity_percolation}{\bf j}), consistent with the rigidity percolation threshold $\rho_r$, suggesting that after the rigidity percolation, most of the links become redundant. This observation shows how each added link affects the rigidity of kirigami - e.g., when $\rho \approx 0.4$, each added link reduces the internal rotational DoF, rather than connecting two connected components, and when $\rho \approx 0.6$, the system is almost connected and rigid because most new links added will be redundant.

\subsection*{Combining two approaches} Deterministic control enables us to achieve certain DoF and NCC precisely based on MRPs and MCPs, while statistical control makes it possible to achieve more combinations of DoF and NCC but only in a random sense. It is natural to ask whether we can combine these two approaches to take advantage of both of them. Given a link density $\rho$, we start with an MRP (or MCP) and randomly add/remove $4L(L-1)|\rho-\rho_r|$ (or $4L(L-1)|\rho-\rho_c|$) links. The experiment is repeated by 100 times, and the average $m_{\text{int}}$, $m_{\text{tot}}$, $T$, and $N$ for different $\rho$ are recorded. 

\begin{figure}[t!]
\centering
\includegraphics[width=0.8\textwidth]{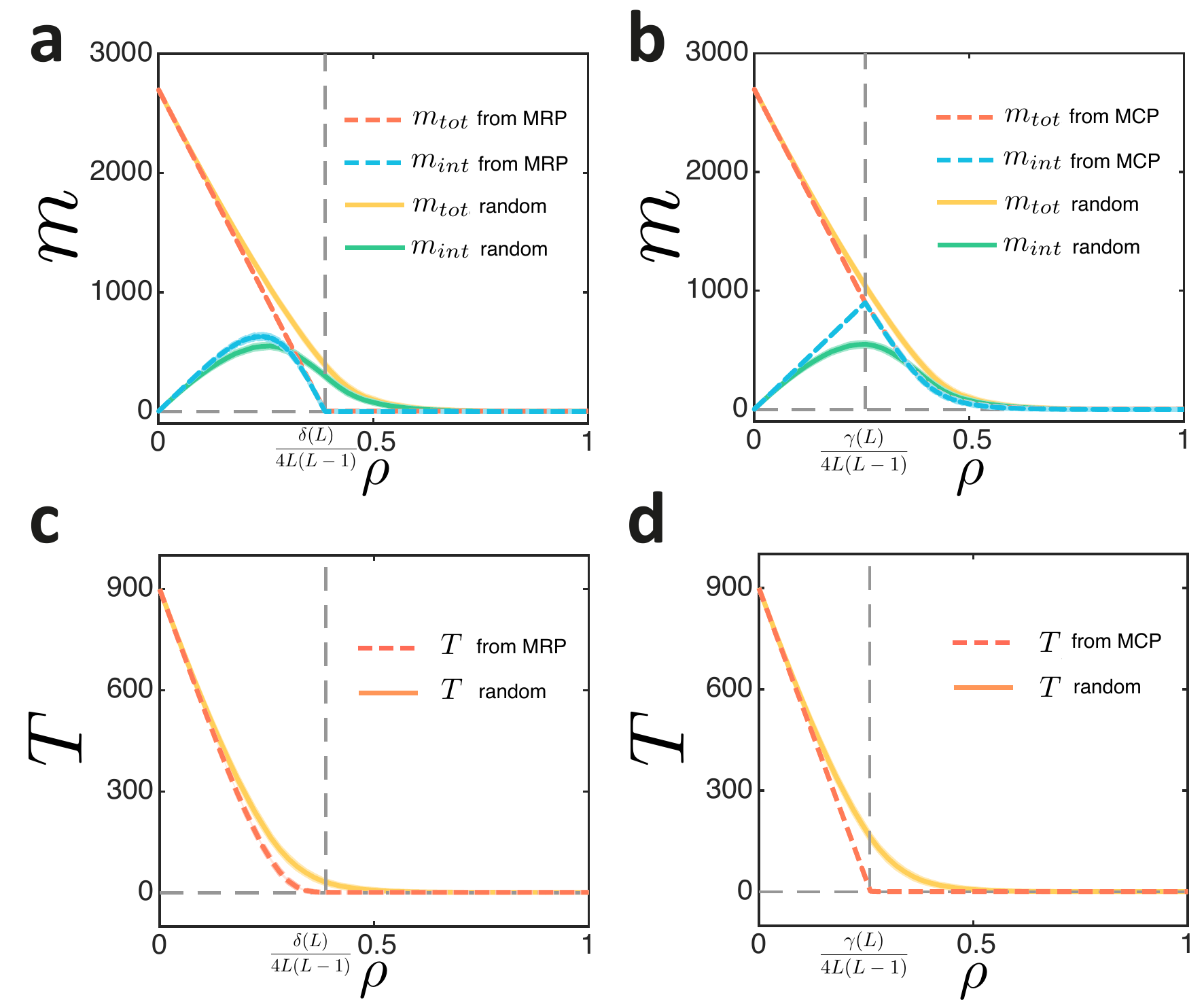}
\caption{{\bf a-b)} The change in the total DoF $m_{\text{tot}}$ and the internal DoF $m_{\text{int}}$ by randomly removing or adding links, starting from {\bf a)} an MRP or a random link pattern, and {\bf b)} an MCP or a random link pattern. {\bf c-d)} The change in the total number of connected components $T$ by randomly removing or adding links, starting from {\bf c)} an MRP or a random link pattern, and {\bf d)} an MCP or a random link pattern.}
\label{fig:random_MRP_MCP}
\end{figure}

For DoF, since all links in MRPs are non-redundant, randomly removing links from them results in a linear change of $m_{\text{tot}}$ (Fig.~\ref{fig:random_MRP_MCP}{\bf a}). More interestingly, starting from an MCP, both adding and removing links randomly result in a decrease of the internal DoF $m_{\text{int}}$ (Fig.~\ref{fig:random_MRP_MCP}{\bf b}), which can be explained by our construction of MCPs (SI Appendix, Fig.~S3). In fact, this is the largest internal DoF that can be achieved. For NCC, $T$ increases gradually as links are removed from MRPs (Fig.~\ref{fig:random_MRP_MCP}{\bf c}). By contrast, a sharp transition of $T$ can be observed as links are removed from MCPs (Fig.~\ref{fig:random_MRP_MCP}{\bf d}). Overall, by combining the deterministic and statistical control, we can achieve a wider range of DoF and NCC, and various behaviors under a perturbation of link density (see SI Appendix, section S4.)

\section*{Discussion}

Our study of rigidity and connectivity in kirigami was made possible by the realization of a one-to-one mapping between the cutting problem and an equivalent linkage problem. This allowed us to provide a bottom-up hierarchical algorithm to construct minimum rigidifying link patterns (MRPs) and minimum connecting link patterns (MCPs), which allow us to rigidify and connect kirigami tessellations optimally. The MRPs and MCPs also provide us with a simple method for obtaining a kirigami system with any given DoF or NCC. Overall this suggests that we can exquisitely control the rigidity and connectivity of kirigami with the topology of prescribed~cuts. 

\begin{figure}[t]
\centering
\includegraphics[width=0.9\textwidth]{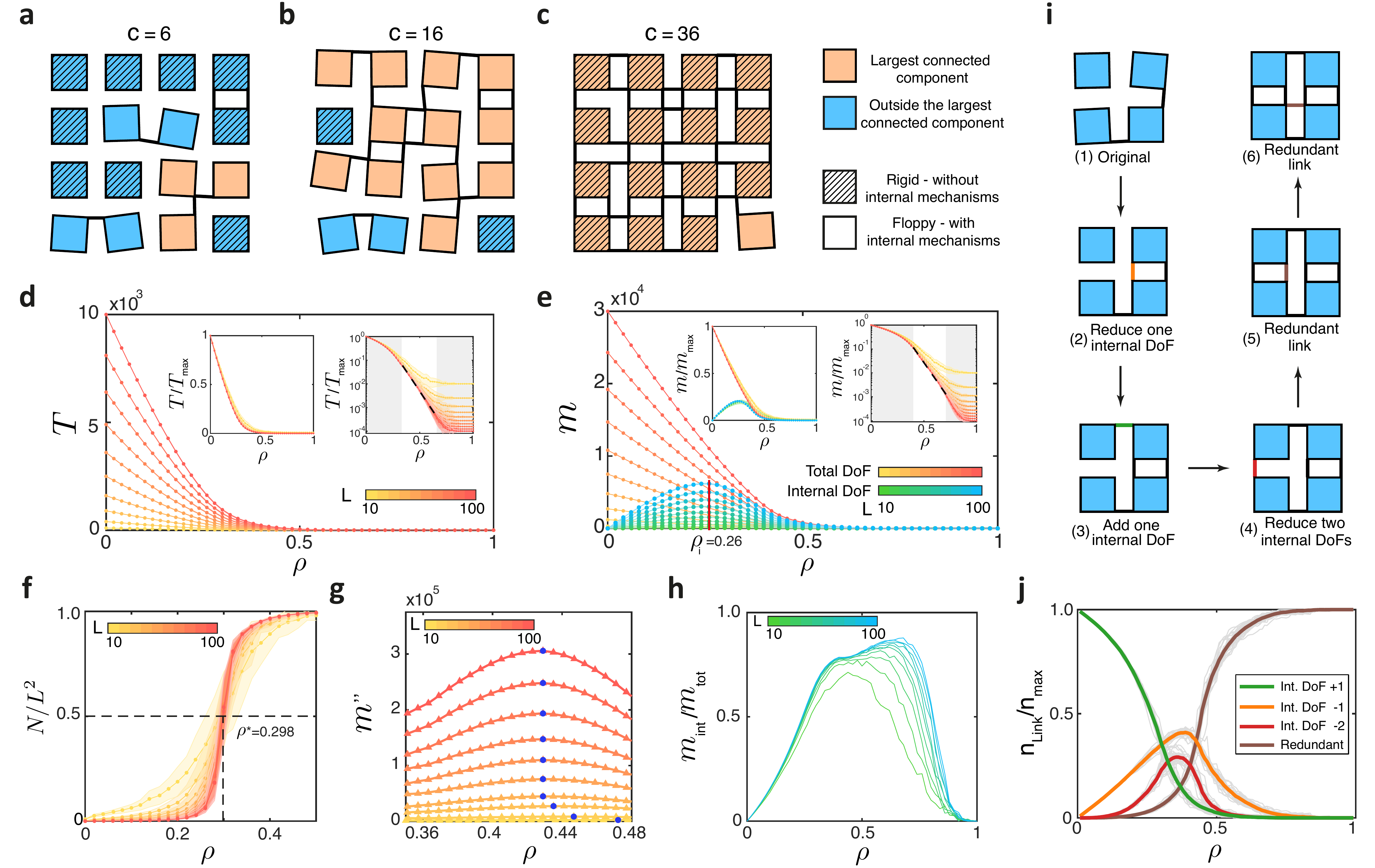}
\caption{Connectedness and rigidity of kirigami patterns.
{\bf a-c)} For an $L\times L=4\times 4$ planar kirigami, links are added randomly. The largest connected component becomes dominant just after a few links are added (orange quads in {\bf b} and {\bf c}). The number of internal modes first increases then decreases. The number of quads with internal mechanisms (not marked by stripes) first increases before decreasing ({\bf a} to {\bf c}). 
{\bf d)} The number of connected components (NCC) $T(\rho)$ decreases as link density $\rho$ increases, first linearly, then exponentially at some intermediate $\rho$ (right inset) before flattening out. This exponential behavior is independent of system size $L$ - if $T$ is scaled by $T_{\text{max}}$ (left inset).
{\bf e)} The change of DoF ($m(\rho)$) shows a similar linear-sublinear transition, exponential decay (right inset) and the scale-independence (left inset) as NCC. The number of internal mechanisms increases, reaches a peak at $\rho_i=0.26$, and approaches the $m_{\text{tot}}$ while decreasing.
{\bf f)} The size of the largest cluster ($N$) has a percolation behavior near $\rho=0.298$. The transition becomes sharper for larger $L$.
{\bf g)} The second derivative of total DoF has a peak at the rigidity percolation threshold. The blue dots represent the peak of each curve, the values of which converge to $0.429$.
{\bf h)} The ratio of the number of internal mechanisms and the number of total mechanisms as a function of $\rho$. The internal mechanism dominates from $\rho\in(0.4,0.8)$. 
{\bf i)} An example showing how the links can affect DoF (rigidity) and $T$ (connectivity) in planar kirigami. From the initial state (1), adding the orange link in state (2) reduces one internal DoF ($\Delta_i=-1, \Delta_r=0$). In state (3), the green links reduces the NCC, but there is one more internal mechanism ($\Delta_i=+1, \Delta_r=-3$). In state (4), the red link freezes the two internal mechanisms, making the system rigid ($\Delta_i=-2, \Delta_r=0$). Finally, in (5) and (6), the link added is redundant. Neither type of DoF is change ($\Delta_i=0, \Delta_r=0$).
{\bf j)} For a system of size $L=30$, as links are added randomly, their influence on DoF is shown in terms of the proportion of each type of link in ({\bf a}) during this process.}
\label{fig:connectivity_percolation}
\end{figure}

At a more coarse-grained level, we also show how to control of connectivity and rigidity by tuning the density of links. We find three critical thresholds: the density for maximum internal DoF ($\rho_i$), for connectivity percolation ($\rho_c$), and for rigidity percolation ($\rho_r$), with $\rho_i <\rho_c<\rho_r$, providing guidance for tuning the link density to achieve different mechanical properties. For example, one can control the ratio of high-frequency and low-frequency modes by choosing the link density above or below the connectivity percolation since the low frequency mode sets in when $\rho >\rho_c$, and the remaining high frequency modes decreases significantly once $\rho > \rho_r$ the rigidity percolation threshold, with relevance for mechanical allostery~\cite{Yan2526, Rocks2520}. Alternatively, since at $\rho_c$ the number of internal modes reaches a maximum, and in light of Fig.~\ref{fig:connectivity_percolation}{\bf j}, it is possible to change the DoF information by removing redundant links and adding links that change the internal DoF, while keeping the total number of links the same; indeed this might allow for click-kirigami (i.e. with reversible links) to be a substrate for mechanical information storage, similar to origami~\cite{Chen19}. 

In a broader context, our theoretical study on the rigidity and connectivity in kirigami complements recent developments in the design and fabrication of physical kirigami structures~\cite{Rafsanjani17, Bertoldi2017,Rafsanjani2018,Sussman2015,Zhang2015} by providing guidelines for their control via internal rotational mechanisms, similar to that seen in origami~\cite{Chen18}. Two natural questions are to investigate functional devices that might exploit this, and to explore the generalization of these arguments to 3D kirigami with polyhedra~\cite{Kadic2019} in the context of architectural and structural design.

{\bf Acknowledgements } This work was supported in part by the Croucher Foundation (to GPTC), National Science Foundation Grant DMR 14-20570 (to LM) and DMREF 15-33985 (to LM).

\end{document}